\documentclass[aps,prl]{revtex4}
\usepackage{graphicx}
\usepackage{subfigure}
\usepackage{amsmath,amsthm}
\usepackage{amssymb,amsfonts}
\usepackage{url}

\begin{document}
	\title{On Thermodynamic Interpretation of Copula Entropy}
	\author{Jian MA}
	\email[Email:]{majian@hitachi.cn}
	\affiliation{Hitachi Research China}
	\begin{abstract}
	Copula Entropy (CE) is a recently introduced concept for measuring correlation/dependence in information theory. In this paper, the theory of CE is introduced and the thermodynamic interpretation of CE is presented with N-particle correlated systems in equilibrium states. 
	\end{abstract}
\maketitle

\section{Introduction}
In the 19th century, thermodynamics was developed, mainly by Clausius\cite{Clausius1854,Clausius1865}, Gibbs\cite{gibbs1902elementary}, and Boltzmann\cite{Boltzmann1872,Boltzmann1877} to describe the phenomenological relationship between the transport of heat to a physical system, work and temperature and therefore the concept of entropy and the second law of thermodynamics were introduced. Boltzmann entropy of a physical system was defined to relate the macrostate of the system to its macrostates as follows
\begin{equation}
\label{eq:be}
	S_B = k\log W,
\end{equation} where $k$ is Boltzmann's constant and $W$ is the number of microstates in equilibrium states. 

Inspired by statistical mechanics, Shannon defined the entropy as measure of information  analogous to thermodynamic entropy in the seminal paper \cite{Shannon1948} when solving the problems of communication. Given random variables $\mathbf{X}$ governed by probability distribution $p(x)$, Shannon's entropy is defined as
\begin{equation}
\label{eq:se}
	H(\mathbf{x})=-K \int_{\mathbf{x}}{p(\mathbf{x})\log{p(\mathbf{x})}d\mathbf{x}},
\end{equation}
where $K$ is a positive constant, usually set as 1. It is the fundamental concept in information theory which is interpreted as the measure of information, choice, or uncertainty. It is essentially the same functional of probability as thermodynamic entropy if $p(\mathbf{x})$ is interpreted as the probability of the system in certain microstates.

Thermodynamic and information theory share intimate historical connections. Among them, Laudauer \cite{Landauer1961} proposed the erasure principle which gives the minimal work required to erase one bit of information on a system and Jaynes \cite{Jaynes1957,Jaynes1957a} proposed the principle of maximum entropy (or MaxEnt for short) for inferring the probability when limited constraints are known. These are two examples which inspire both physicsists and information theorists to build more connections between the two theories.

Mutual Information (MI) is another fundamental concept in information theory \cite{cover1999elements}, which is defined for measuring the information of one variables $X$ contained in another variable $Y$ as follows:
\begin{equation}
\label{eq:mi}
	I(X,Y) = H(X) - H(X|Y).
\end{equation}
Since MI is a bivariate measure, researchers have contributed much works to find its multivariate extensions. In 2008, Ma and Sun defined a new concept, called Copula Entropy (CE), and proved its equivalence to MI \cite{Ma2011}. CE is defined with copula theory\cite{nelsen2007introduction} which is the theory about representing the coupling of random variables with a copula function. In this sense, CE is essentially a Shannon entropy defined on unit space for measuring such couplings in copula functions. It has several ideal properties, such as symmetric, multivariate, invariant to monotonic transformation, etc. Furthermore, it is proved that Shannon entropy of random variables can be decomposed into CE and the sum of marginal entropies of each variable. CE has also theoretical relationship  \cite{jian2019estimating} with another important information theoretical measure for causality -- Transfer Entropy. Since its birth, CE has been widely applied to many different disciplines, including hydrology\cite{chen2019copulas,chen2013measure}, biology\cite{charzynska2015improvement}, neuroscience\cite{ince2016the,kayser2015irregular},medicine\cite{Mesiar2021}, among many others.

As a rigorous defined Shannon entropy, CE has clear mathematical interpretation. However, its thermodynamic interpretation has been absent yet. Physicists may be interested in what CE means in physical systems, just as Shannon entropy has its counterpart in thermodynamics. What is CE, the entropy of coupling, in thermodynamic systems? How thermodynamic entropy can be decomposed just as Shannon entropy does into marginal entropies and CE? What kind of role CE can play in thermodynamics?

In this paper, we will present a thermodynamic interpretation of CE as the entropy of interaction in N-particle systems, such as dense gas or liquid. Thermodynamic entropy of the system will be decomposed into the entropy of non-interacting particles and the correlation entropy due to interactions, i.e., CE. Such interpretation is potentially applicable to many problems, such as calculation of entropy of liquid water \cite{Lazaridis1996,Baranyai1989,Wallace1987,Baranyai1991}, liquid metals\cite{Gao2018b}, or others\cite{Tripodo2021}.

\section{Theory of Copula Entropy}
CE is the theory of measure of statistical correlation/dependence and has its root in the copula theory -- the theory of representation of correlation/dependence between random variables \cite{nelsen2007introduction}. It is defined with copula function in copula theory. According to Sklar's theorem\cite{sklar1959fonctions}, given random variables $\mathbf{X}=\{X_1,\ldots,X_N\}$ and theirs marginal functions $\mathbf{u}=\{u_1,\ldots,u_N\}$, the joint probability density function $p(\mathbf{x})$ can be representation as follows:
\begin{equation}
\label{eq:sklar}
	p(\mathbf{x})=c(\mathbf{u})\prod_{i}{p(x_i)},
\end{equation}
where $c(\cdot)$ is the copula density function which contains all the dependent information between random variables $\mathbf{x}$. The theorem gives a universal formulation which separates the coupling of $\mathbf{x}$ from individual variables. When $\mathbf{X}$ are mutually independent, $c(\mathbf{u})=1$.

With this representation, one can define CE in the same formula as Shannon entropy just by replacing probability function $p(\mathbf{x})$ with copula density function $c(\mathbf{u})$, as follows:
\begin{equation}
\label{eq:ce}
	H_c(\mathbf{x}) = -\int_{\mathbf{u}}c(\mathbf{u})\log{c(\mathbf{u})d\mathbf{u}}.
\end{equation}
It is easy to understand that $H_c(\mathbf{x})$ measures the information in copula function $c(\mathbf{u})$. Since $c(\mathbf{u})$ is a probability density function on unit space and $H_c$ is also a kind of Shannon entropy. Ma and Sun\cite{Ma2011} proved that CE is equivalent to negative MI:
\begin{equation}
	I(\mathbf{x})=-H_c(\mathbf{x}).
\end{equation} 
Here, CE is actually multivariate MI and can characterize the independence between random variables since $H_c(\mathbf{x})=0$ in the cases of independence. Combining \eqref{eq:se}, \eqref{eq:sklar} and \eqref{eq:ce}, one can derive another theorem which states the relationship between Shannon entropy, marginal entropies and CE:
\begin{equation}
\label{eq:co}
	H(\mathbf{x})=\sum_{i}{H(x_i)}+H_c(\mathbf{x}).
\end{equation}
This theorem means that Shannon entropy can be decomposed into two independent parts: marginal entropies for individual variables and CE for the coupling between random variables.

As is well known, Shannon entropy can be derived from the four Khinchin-Shannon axioms\cite{Shannon1948}: continuity, maximality, expansibility, and separability (or strong additivity). Particular, the axiom of separability states that the entropies of two independent subsystems equals that of the system composed of two subsystems. The \eqref{eq:co} generalizes the axioms of separability to more general cases of dependent subsystems. This makes it possible to build a framework of axioms for deriving thermodynamics \cite{Lieb1999,Lieb2013,Lieb2014,Weilenmann2016} for cases, including interacting and non-interacting systems, equilibrium and non-equilibrium states.

\section{Thermodynamic interpretation}
\subsection{System}
Let us consider a N-particle thermodynamic system in equilibrium states with interaction between systems. The particles have mass $m$, position $\mathbf{q}_i$, and momenta $\mathbf{p}_i$. The Hamiltonian $H_N$ of each N-particle system is given by
\begin{equation}
\label{eq:hn}
	H_N = \sum_{i=1}^{N}{\frac{\mathbf{p}_i^2}{2m}}+V_N,
\end{equation}
where $V_N$ is the total potential of the system. At given temperature $T$, the canonical probability function $P_N$ is 
\begin{equation}
\label{eq:pn}
	P_N = \frac{e^{-\beta H_N}}{Z_N},
\end{equation}
where $\beta = (k_B T)^{-1}$, $T$ is temperature, and $Z_N$ is the partition function as
\begin{equation}
\label{eq:zn}
	Z_N=\frac{1}{h^{3N}N!}\int \cdots \int {e^{-\beta H_N}d\mathbf{q}_1 d\mathbf{p}_1\cdots d\mathbf{q}_N d\mathbf{p}_N}.
\end{equation}
Then the total canonical entropy $S_N$ is
\begin{equation}
\label{eq:sn}
	S_N = -\frac{k_B}{h^{3N}N!}\int {P_N\ln P_N d\mathbf{q}_1 d\mathbf{p}_1\cdots d\mathbf{q}_N d\mathbf{p}_N}.
\end{equation}

the canonical probability function in \eqref{eq:pn} can be expressed as
\begin{equation}
\label{eq:fnn1}
	f_N^{(N)}(\mathbf{q}_1,\mathbf{p}_1,\ldots,\mathbf{q}_N,\mathbf{p}_N)=f_N^{(1)}(\mathbf{p}_1)\cdots f_N^{(1)}(\mathbf{p}_N)g_N^{(N)}(\mathbf{q}_1,\ldots,\mathbf{q}_N),
\end{equation}
where $f_N^{(1)}$ is the probability density of one particle and $g_N^{(N)}$ is the N-particle correlation function. For $n$-particle correlated systems, the canonical probability function similar to \eqref{eq:fnn1} is 
\begin{equation}
\label{eq:fnn2}
		f_N^{(n)}(\mathbf{q}_1,\mathbf{p}_1,\ldots,\mathbf{q}_n,\mathbf{p}_n)=f_N^{(1)}(\mathbf{p}_1)\cdots f_N^{(1)}(\mathbf{p}_n)g_N^{(n)}(\mathbf{q}_1,\ldots,\mathbf{q}_n),
\end{equation}
where $g_N^{(n)}$ is derived by integrating $g_N^{(N)}$ over $\mathbf{q_{n+1}},\ldots,\mathbf{q}_N$. Then the total canonical entropy $S_N$ in \eqref{eq:sn} is re-expressed in terms of $f_N^{(N)}$ as
\begin{equation}
\label{eq:sn2}
	S_N = -\frac{k_B}{N!}\int{\prod_{i}{d \mathbf{q}_i d \mathbf{p}_i f_N^{(N)} \ln(h^{3}f_N^{(N)})}}.
\end{equation}
Combining \eqref{eq:fnn1} and \eqref{eq:sn2}, one can derive the representation of total canonical entropy $S_N^{(N)}$ as
\begin{equation}
\label{eq:sn1g}
	S_N = S_N^{(1)} + S_N^{g},
\end{equation}
where $S_N^{(1)}$ are the entropies of one particle and $S_N^{g}$ is the entropy of correlation function defined as
\begin{align}
\label{eq:sn1}
	S_N^{(1)}&=-\frac{k_B}{N!} \int{f_N^{(1)}\ln{h^3}f_N^{(1)}(\mathbf{p}_1)d\mathbf{p}_1}\\
	S_N^{g}&=-\frac{k_B}{N!} \int{g_N^{(N)}\ln{h^3}g_N^{(N)}d\mathbf{q}_1}\ldots d\mathbf{q}_N
\end{align}
The entropy of correlation function $S_N^g$ can further be expanded into a series in terms of high order correlations \cite{green1969molecular} as
\begin{align}
\label{eq:sng}
	S_N^g =&-\frac{1}{2!}k_B \rho_N^2\int{g_N^{(2)}(\mathbf{q}_1,\mathbf{q}_2)\ln{g_N^{(2)}(\mathbf{q}_1,\mathbf{q}_2)}d\mathbf{q}_1 d\mathbf{q}_2}\\
	&-\frac{1}{3!}k_B \rho_N^3\int{g_N^{(3)}(\mathbf{q}_1,\mathbf{q}_2,\mathbf{q}_3)\ln{\delta g_N^{(3)}(\mathbf{q}_1,\mathbf{q}_2,\mathbf{q}_3)}d\mathbf{q}_1 d\mathbf{q}_2 d\mathbf{q}_3}\\
	&\cdots
\end{align}
where $\rho_N=N/V$ and $\delta g_N^{(n)}$ is the high order correlation function in terms of low order functions $g_N^{(n-1)}$ by enumerating all possible combinations.

\subsection{Interpretation}
Here we present the thermodynamic interpretation of CE. The representation of the canonical probability function in \eqref{eq:fnn1} separates the probability of one-particle from the N-particle correlation function, which actually plays the same role as Sklar's theorem \eqref{eq:sklar} does in copula theory. According to Sklar's theorem \eqref{eq:sklar}, the correlation function $g_N^{(N)}$ is the copula function which contains all the information of the interaction between systems and is separated from the probability of one-particle $f_N^{(1)}$. So, we can interpret copula function as the representation of the coupling between particles in this interactive systems. Copula functions represent all the coupling information in one function, including all order interactions between particles, both local and non-local. Copula function provides a universal theoretical representation of the interactive systems.

With the representation of copula function, one can interpret the results \eqref{eq:sn1g} according to the theorem \eqref{eq:co}, which decomposes the total canonical entropy $S_N$ into the part $S_N^{(1)}$ for the entropies of each particle and the entropy of the coupling of particles $S_N^g$. The coupling entropy $S_N^g$ defined with $g_N^{(N)}$ measures the information containing in the correlation function. So the thermodynamic meaning of CE is the entropy of N-particle correlation due to interacting particles. 

In the uncorrelated cases of ideal non-interactive particle systems, no coupling exists and therefore $g_N^{(N)}=1$ and $S_N^g=0$ because the canonical probability function can be expressed with only $f_N^{(1)}$. This corresponds to the cases of independent random variables where the joint probability function is the product of marginal functions, and hence copula function $c(\mathbf{u})=1$ and CE equals 0.

In the correlated cases, copula function contains all the information about interacting particles, both local and long range, low order and high order. Therefore CE or $S_N^g$ measures the exact information of interacting particles, including long range correlation, and all order of correlations. The entropy of correlation function $S_N^g$ can be expressed as a series of high order correlations in \eqref{eq:sng}. It is also easy to know that CE can also be expressed as a series of high order moment functions. If one can derive the copula function for interactive systems, CE makes it possible to calculate the entropy of the systems, like dense fluids, without making approximations by eliminating non-locality terms \cite{Wallace1987,Baranyai1989,Gao2018b,Laird1992,Yokoyama2001}.

CE is non-positive because correlation between random variables leads to random variables contain the information of other variables and therefore the total amount of information of joint distribution is reduced and is smaller than the sum of marginal entropies. In thermodynamics, this can be interpreted as the reduction of the total entropy of the systems due to interacting or the reduced uncertainty of microstates of thermodynamic systems because the coupling particle will make more certainty of the microstates. In this sense, CE means how much works can be extracted from N-particle correlated systems by strengthening the coupling of particles. Since copula function is independent of marginal functions and joint functions, this part of works are also independent of total works that can be extracted from the systems. That also implies that one can extract works from systems by simply enhancing the coupling of particles as indicated by the following equation
\begin{equation}
\label{eq:cfe}
W = E - T\Delta S.
\end{equation}
where $W,E,T$ and $\Delta S$ are work, energy, temperature and the change of entropy due to coupling respectively.

CE is a rigourous mathematical concept for any correlation and is universal applicable. So, it can be interpreted not only for equilibrium but non-equilibrium states. Hence we can use it to study entropy production in non-equilibrium states. In non-equilibrium states, the probability of microstate changes and the coupling of particles changes also. As the representation of coupling particles, copula function may change from one function family to another and hence CE may change as consequence. The change of CE is a part of entropy production. If only the coupling of particles changes, then the change of CE will be all of entropy production.

\bibliographystyle{unsrt}
\bibliography{thermo1}

\end{document}